\documentclass[prx, preprint, superscriptaddress]{revtex4-1}

\usepackage{graphicx}  
\usepackage{bm} 
\usepackage{amssymb}  
\usepackage{amsmath}
\usepackage{mhchem}
\usepackage{mathptmx}
\usepackage{siunitx}
\usepackage{afterpage}
\usepackage{color}
\begin{document}
\widetext


\title{Observation of the polaronic character of excitons in a two-dimensional semiconducting magnet \ce{CrI3}}


\author{Wencan Jin}
\altaffiliation{current affiliation: Department of Physics, Auburn University, 380 Duncan Drive, Auburn, AL 36849, USA}
\affiliation{Department of Physics, University of Michigan, 450 Church Street, Ann Arbor, Michigan 48109, USA}

\author{Hyun Ho Kim}
\altaffiliation{current affiliation: School of Materials Science and Engineering, Kumoh National Institute of Technology, Gumi, Gyeongbuk 39177, Korea}
\affiliation{Institute for Quantum Computing, Department of Chemistry, and Department of Physics and Astronomy, University of Waterloo, Waterloo, 200 University Ave W, Ontario N2L 3G1, Canada}

\author{Zhipeng Ye}
\affiliation{Department of Electrical and Computer Engineering, 910 Boston Avenue, Texas Tech University, Lubbock, Texas 79409, USA}

\author{Gaihua Ye}
\affiliation{Department of Electrical and Computer Engineering, 910 Boston Avenue, Texas Tech University, Lubbock, Texas 79409, USA}

\author{Laura Rojas}
\affiliation{Department of Electrical and Computer Engineering, 910 Boston Avenue, Texas Tech University, Lubbock, Texas 79409, USA}

\author{Xiangpeng Luo}
\affiliation{Department of Physics, University of Michigan, 450 Church Street, Ann Arbor, Michigan 48109, USA}

\author{Bowen Yang}
\affiliation{Institute for Quantum Computing, Department of Chemistry, and Department of Physics and Astronomy, University of Waterloo, Waterloo, 200 University Ave W, Ontario N2L 3G1, Canada}

\author{Fangzhou Yin}
\affiliation{Institute for Quantum Computing, Department of Chemistry, and Department of Physics and Astronomy, University of Waterloo, Waterloo, 200 University Ave W, Ontario N2L 3G1, Canada}

\author{Jason Shih An Horng}
\affiliation{Department of Physics, University of Michigan, 450 Church Street, Ann Arbor, Michigan 48109, USA}

\author{Shangjie Tian}
\affiliation{Department of Physics and Beijing Key Laboratory of Opto-electronic Functional Materials \& Micro-nano Devices, Renmin University of China, Beijing 100872 China}

\author{Yang Fu}
\affiliation{Department of Physics and Beijing Key Laboratory of Opto-electronic Functional Materials \& Micro-nano Devices, Renmin University of China, Beijing 100872 China}

\author{Hui Deng}
\affiliation{Department of Physics, University of Michigan, 450 Church Street, Ann Arbor, Michigan 48109, USA}

\author{Hechang Lei}
\affiliation{Department of Physics and Beijing Key Laboratory of Opto-electronic Functional Materials \& Micro-nano Devices, Renmin University of China, Beijing 100872 China}

\author{Kai Sun}
\affiliation{Department of Physics, University of Michigan, 450 Church Street, Ann Arbor, Michigan 48109, USA}

\author{Adam W. Tsen}
\affiliation{Institute for Quantum Computing, Department of Chemistry, and Department of Physics and Astronomy, University of Waterloo, Waterloo, 200 University Ave W, Ontario N2L 3G1, Canada}

\author{Rui He}
\email{rui.he@ttu.edu}
\affiliation{Department of Electrical and Computer Engineering, 910 Boston Avenue, Texas Tech University, Lubbock, Texas 79409, USA}

\author{Liuyan Zhao} 
\email{lyzhao@umich.edu}
\affiliation{Department of Physics, University of Michigan, 450 Church Street, Ann Arbor, Michigan 48109, USA}



\begin{abstract}
\noindent Exciton dynamics can be strongly affected by lattice vibrations through electron-phonon (e-ph) coupling, forming polarons \cite{RN1} -- composite quasiparticles with electronic states dressed by a cloud of polar phonons. In two-dimension (2D) atomic crystals, polaronic character of excitons has only been studied in a couple of transition metal dichalcogenide (TMDC) semiconductors \cite{RN2, RN3, RN4, RN5}, yet has already revealed its importance in determining the physical properties of the host materials \cite{RN2, RN3, RN4, RN5}. The recently discovered 2D magnetic semiconductor, \ce{CrI3}, provides another new platform to explore polaronic character of excitons beyond the classic 2D TMDC semiconductors because of its non-Wannier excitons \cite{RN6, RN7} and  intrinsic long-range magnetic order \cite{RN8}. Here, focusing on bilayer \ce{CrI3}, we first performed temperature-dependent photoluminescence and linear absorption measurements to show the presence of strong e-ph coupling. We then carried out careful temperature and magnetic field-dependent resonant micro-Raman measurements and discovered a rare Raman feature of periodic broad modes up to the $8^\mathrm{th}$ order (well beyond the multiphonon order of $3^\mathrm{rd}$), which can be attributed to the polaronic character of excitons. We establish that the polaronic character in bilayer \ce{CrI3} is dominated by the coupling between the bright exciton at 1.96 eV and a longitudinal optical phonon at 120.6 $\mathrm{cm}^{-1}$. We further show that the emergence of long-range magnetic order below $T_\mathrm{C}$ = 45 K enhances the e-ph coupling strength by nearly 50\% and that the transition from layered antiferromagnetic to ferromagnetic order across $B_\mathrm{C}$ = 0.7 T tunes the spectral intensity of the periodic broad modes, suggesting a strong coupling among the lattice, charge and spin degrees of freedom in 2D \ce{CrI3}. Our study opens up new avenues for tailoring light-matter interactions in 2D semiconductors with long-range orders.
\end{abstract}
\maketitle


The polaronic effect, which describes the strong coupling between charge and lattice vibrations, plays a key role in a broad class of novel quantum phenomena ranging from colossal magnetoresistance \cite{RN9} to anomalous photovoltaic effect \cite{RN10}. In particular, the polaronic effect on excitons can profoundly modulate exciton dynamics upon photoexcitation and has been employed to describe intriguing optical and optoelectronic properties in materials such as hybrid organic-inorganic perovskite solar cells \cite{RN11,RN12,RN13}. Compared to three-dimensional (3D) bulk systems, two-dimensional (2D) atomic crystals possess a couple of unique advantages in exploring the polaronic effect on exciton dynamics. First, the reduced dielectric screening in atomically thin samples enhances both the excitonic effect \cite{RN14} and the electron-phonon (e-ph) coupling \cite{RN15}, which is expected to promote the polaronic effect of excitons. Second, unlike bulk materials in which the e-ph coupling is largely determined by intrinsic electronic and phonon band structures with limited tunability, 2D materials provide greater flexibility for engineering e-ph coupling through a battery of approaches including carrier doping \cite{RN2, RN3} and interfacial coupling \cite{RN4, RN5, RN16, RN17, RN18}, as well as dimensionality modulation \cite{RN15}, and therefore hold high promise for the future development of optoelectronic devices.
 
The realization of a long-range magnetic order in 2D semiconducting \ce{CrI3} paves the way to engineer optical and optoelectronic properties of 2D semiconductors \cite{RN19, RN20, RN21, RN22, RN23, RN24, RN25, RN26}. The large excitonic effect \cite{RN7} from the localized molecular orbitals, of neither Wannier-type in 2D TMDCs nor Frenkel-type in ionic crystals, can be considered as the microscopic origin of the giant magneto-optical Kerr effect \cite{RN8} and magnetic circular dichroism \cite{RN6} signals in 2D \ce{CrI3}. Meanwhile, the strong e-ph coupling is suggested to cause the large Stokes shift, profound broadness, and skewed lineshape in the photoluminescence (PL) spectra of 2D \ce{CrI3} \cite{RN6}. The coexistence of excitons and strong e-ph coupling in 2D \ce{CrI3} naturally leads to open experimental questions of whether polaronic character emerges in the exciton dynamics and whether they are affected by the long-range magnetic order.

One fingerprint for polaronic effect is the development of phonon-dressed electronic bands that appear as satellite bands in proximity to the original undressed one. Such features manifest as multiple equally spaced replica bands in angle-resolved photoemission spectroscopy (ARPES) \cite{RN2, RN17, RN27, RN28, RN29, RN30, RN31, RN32} or as discrete absorption and emission lines in linear optical spectroscopy \cite{RN33}. However, such signatures of polaronic effect have not been revealed so far in 2D \ce{CrI3}, as the sizable bandgap ($\sim$1.1 eV) \cite{RN6} and extreme surface sensitivity \cite{RN34} of \ce{CrI3} make ARPES measurement challenging while the potential inhomogeneous broadening could largely smear out individual lines for phonon-dressed bands in linear optical spectroscopy. In this work, we exploit temperature and magnetic field dependent resonant micro-Raman spectroscopy, to show the direct observation of the polaronic character of excitons in bilayer \ce{CrI3}. The polaronic effect manifests in the Raman spectra as a well-defined, periodic pattern of broad modes that is distinct from sharper phonon peaks. The profile of this periodic pattern and its temperature and magnetic field dependence reveal essential information including the e-ph coupling strength and the tunability of polaronic effect by the magnetism in bilayer \ce{CrI3}. We mainly focus on bilayer \ce{CrI3} because it features a single magnetic phase transition from the layered antiferromagnetic (AFM) to ferromagnetic (FM) order and briefly compare to results on thicker \ce{CrI3} flakes afterwards.

We start by identifying excitonic transitions and e-ph coupling in bilayer \ce{CrI3} using temperature dependent PL and linear absorption spectroscopy. Bilayer \ce{CrI3} was fully encapsulated between few-layer hexagonal boron nitride (hBN) and placed on a sapphire substrate (see Methods). Photoluminescence and linear absorption spectroscopy measurements were then performed in a transmission geometry (see Methods). Figure 1 shows PL and absorbance spectra taken at 80 K, 40 K, and 10 K that are well above, slightly below, and well below the magnetic critical temperature $T_\mathrm{C}$ = 45 K, respectively \cite{RN6, RN8, RN34}. A single PL mode at 1.11 eV and three prominent absorbance peaks at 1.51 eV, 1.96 eV, and 2.68 eV (denoted as A, B, and C, respectively) are observed across the entire temperature range. These three energies are in good agreement with the ligand-field electronic transitions assigned by differential reflectance measurements on monolayer \ce{CrI3} \cite{RN6} and bulk \ce{CrI3} \cite{RN35, RN36}, and have been later revealed to be bright exciton states through more recent sophisticated first principle GW and Bethe-Salpeter equation (GW-BSE) calculations \cite{RN7}. The large Stokes shift ($\sim$400 meV) between the PL and A exciton absorption peak is consistent with previous report \cite{RN6} and indicates strong e-ph coupling in 2D \ce{CrI3}. While the absorbance spectra show little temperature dependence except the appearance of a weak shoulder at 1.79 eV at 10 K (orange arrow), the PL spectra are clearly temperature dependent. In particular, the temperature dependence of the PL full width at half maximum, $\varGamma(T)$, is well fitted by the model functional form, $\varGamma(T) = \varGamma_0 + \frac{\gamma}{\mathrm{exp}(\frac{\hbar\omega_\mathrm{LO}}{{k_B}T})-1}$, with the first term for temperature-independent inhomogeneous broadening and the second term for homogeneous broadening from exciton coupled with a longitudinal optical (LO) phonon at frequency $\omega_\mathrm{LO}$. Taking $\omega_\mathrm{LO}$ = 120.6 $\mathrm{cm}^{-1}$ found later on in Fig. 2, we obtain $\varGamma_0$ = 163.9 $\pm$ 2.7 meV and $\gamma$ = 164.2 $\pm$ 8.1 meV, which suggests that the broadness of the exciton modes arises from both inhomogeneous broadening from impurities or disorders and e-ph coupling. The large homogeneous broadening parameter ($\gamma$) indicates strong vibronic modes mixing in the PL spectra which precludes the formation of well-resolved phonon sidebands \cite{RN6}.

We next proceed to perform resonant micro-Raman spectroscopy measurements with an incident wavelength of 633 nm matching the energy of the B exciton on an encapsulated bilayer \ce{CrI3} flake placed on a \ce{SiO2}/Si substrate (see Methods). Figure 2a displays a representative Raman spectrum acquired in the crossed linear polarization channel at 40 K (slightly below $T_\mathrm{C}$ = 45 K). Note that this spectrum covers a much wider frequency range than earlier Raman studies on \ce{CrI3} \cite{RN34, RN37, RN38, RN39, RN40, RN41, RN42, RN43, RN44}. The multiphonon scattering is visible up to the $3^\mathrm{rd}$ order, and their zoom-in Raman spectra are shown in the inset of Fig. 2a. The $1^\mathrm{st}$-order single phonon peaks appear in the relatively low frequency range of 50 -- 150 $\mathrm{cm}^{-1}$, and are assigned to be of either $A_\mathrm{g}$ or $E_\mathrm{g}$ symmetries under the $C_\mathrm{3i}$ point group (Supplementary Section 1), which is consistent with earlier work \cite{RN34, RN37, RN38, RN39, RN40, RN41, RN42, RN43, RN44} and proves the high quality of our samples. The $2^\mathrm{nd}$-order two-phonon and $3^\mathrm{rd}$-order three-phonon modes show up in slightly higher frequency range of 190 -- 290 $\mathrm{cm}^{-1}$ and 310 -- 410 $\mathrm{cm}^{-1}$, respectively, and show decreasing mode intensities at higher order processes, same as typical multiphonon overtones under harmonic approximation \cite{RN45} or cascade model \cite{RN46}. In addition to and distinct from these multiphonon features, we resolve a remarkable periodic modulation across a wide frequency range of 70 -- 800 $\mathrm{cm}^{-1}$ in the low intensity part of the Raman spectrum (highlighted by the orange shaded area in Fig. 2a). This low intensity periodic pattern consists of clean, individual Lorentzian profiles and survives up to the $8^\mathrm{th}$ order (Fig. 2b), well beyond the highest order ($3^\mathrm{rd}$-order) of multiphonon overtones, and each order of it spans for nearly 50 $\mathrm{cm}^{-1}$ frequency range, much wider than the linewidth of any observed phonon modes (insets of Fig. 2a for phonons). Such a periodic pattern is also observed in the anti-Stoke's side at higher temperatures in bilayer \ce{CrI3}, for example, up to the -$2^\mathrm{nd}$ order at 200 K (Supplementary Section 2), which clearly supports its Raman origin instead of luminescence.

We fit this low intensity periodic pattern using a summation of Lorentzian profiles of the form $\sum_N\frac{A_N(\varGamma_N/2)^2}{(\omega-{\omega}_N)^2+(\varGamma_N/2)^2}+C$ with central frequency $\omega_N$, linewidth $\varGamma_N$, peak intensity $A_N$ of the $N^\mathrm{th}$ period, and a constant background $C$ (see fitting procedure in Methods).  Among all eight orders ($N = 1, 2, \cdots ,8$) in Fig. 2b, the presence of the $1^\mathrm{st}$-order broad mode is deliberately validated in Fig. 2c that fitting with this $1^\mathrm{st}$-order broad mode (orange shaded broad peak in the bottom panel) is visibly better than without it (top panel). This improved fitting by involving the $1^\mathrm{st}$ order broad mode is further rigorously confirmed by the bootstrap method \cite{RN47} (Supplementary Section 3). Figure 2d shows the plot of the central frequency $\omega_\mathrm{N}$ as a function of $N$ with data taken at 40 K ($N = 1, 2, \cdots ,8$) and 290 K ($N = -2, -1, \cdots ,3$), from which a linear regression fit gives a periodicity of 120.6 $\pm$ 0.9 $\mathrm{cm}^{-1}$ and an interception of 0 $\pm$ 0.2 $\mathrm{cm}^{-1}$. 

To the best of our knowledge, such a periodic pattern made of individual Lorentzian profiles previously has only been seen in multiphonon Raman spectra of Cd, Yb, and Eu monochalcogenides described by configuration-coordinate model \cite{RN48, RN49, RN50, RN51, RN52, RN53, RN54, RN55}. However, the periodic pattern observed in bilayer \ce{CrI3} here differs from these monochalcogenide multiphonon modes, as the broad linewidth of $1^\mathrm{st}$-order mode contradicts with the sharp $1^\mathrm{st}$-order forbidden LO-phonon in Cd and Yb monochalcogenides \cite{RN48, RN49, RN50, RN51} and the persistence (or even enhancement) of higher-order multiphonon below $T_\mathrm{C}$ = 45 K is in stark contrast to the disappearance of paramagnetic spin disorder-induced multiphonon below magnetic phase transitions in Eu monochalcogenides \cite{RN52, RN53, RN54, RN55}. Because no known multiphonon model can capture all characteristics of our observed periodic pattern as well as the broad linewidths of each mode, we are inspired to consider the electronic origin. Indeed, strikingly similar features have been seen in polaron systems through the energy dispersion curves (EDCs) of ARPES \cite{RN2, RN17, RN27, RN28, RN29, RN30, RN31, RN32} and linear absorption and PL spectroscopy \cite{RN12, RN33, RN56}. In those cases, the periodic patterns in their energy spectra were determined as the phonon-dressed electronic state replicas, or sometimes also referred as phonon-Floquet states \cite{RN57}, and the periodicity is given by the frequency of the coupled phonons. Due to the high resemblance between the lineshapes of our Raman spectrum and those polaron energy spectra \cite{RN12, RN17, RN27, RN28, RN29, RN30, RN31, RN32, RN56}, we propose that this periodic pattern in Raman spectra of 2D \ce{CrI3} stems from inelastic light scattering between the phonon-dressed electronic states caused by the polaronic character of B excitons in 2D \ce{CrI3}, whereby the B exciton at 1.96 eV, with the electron (hole) in the weakly dispersive conduction (highly dispersive valence) band of Cr $3d$ (I $5p$) orbital character \cite{RN7, RN58}, couples strongly to a phonon at 120.6 $\mathrm{cm}^{-1}$ \cite{RN59}. It is worth noting that a recent theoretical work predicts magnetic polaronic states in 2D \ce{CrI3} because of charge-magnetism coupling \cite{RN60} whereas our work suggests polaronic exciton states due to charge-lattice coupling. 

We then proceed to identify the source and the character of the phonon at 120.6 $\mathrm{cm}^{-1}$. We first rule out the possibility of this phonon arising from either the hBN encapsulation layers or the \ce{SiO2}/Si substrate, as a similar periodic pattern in the Raman spectrum is also observed in bare bulk \ce{CrI3} crystals (see Supplementary Section 4). Compared to the calculated phonon band dispersion of monolayer \ce{CrI3} \cite{RN61}, we then propose the LO phonon calculated to be at about 115 $\mathrm{cm}^{-1}$ as a promising candidate, whose slight energy difference from the experimental value of 120.6 $\mathrm{cm}^{-1}$ could result from calculation uncertainties. This LO phonon mode belongs to the parity-odd $E_\mathrm{u}$ symmetry of the $C_\mathrm{3i}$ point group, and its atomic displacement transforms like an in-plane electronic field ($E_x$, $E_y$) (see inset of Fig. 2d) \cite{RN61}. Its odd parity makes it Raman-inactive and absent in the $1^\mathrm{st}$-order phonon spectra (Fig. 2a inset, top panel), whereas its polar displacement field allows for its strong coupling to electrons/holes and prompts the polaronic character of the charge transfer B exciton (see Supplementary Section 5 for measurements with additional laser wavelengths). In addition, this LO phonon band is nearly dispersionless and has a large density of states, further increasing its potential for coupling with the B excitons in 2D \ce{CrI3}.

Given the coexistence of a 2D long-range magnetic order and polaronic effect of excitons below $T_\mathrm{C}$ = 45 K in bilayer \ce{CrI3}, it is natural to explore the interplay between the two. For this, we have performed careful temperature-dependent Raman spectroscopy measurements and fitted the periodic pattern in every spectrum with a sum of Lorentzian profiles. Figure 3a displays the periodic pattern in Raman spectra taken at 70 K and 10 K, well above and below $T_\mathrm{C}$, respectively. Comparing these spectra, not only do more high-order replica bands become visible at lower temperatures (\textit{i.e.}, from $N$ = 6 at 70 K to $N$ = 8 at 10 K), but also the spectral weight shifts towards the higher-order bands (\textit{i.e.}, from $N$ = 1 at 70 K to between $N$ = 3 and 4 at 10 K in Fig. 3b). The appearance of higher-order modes at lower temperatures possibly results from a combination of the narrow exciton linewidth (about 50 $\mathrm{cm}^{-1}$) and the dispersionless nature of coupled LO phonon. More importantly, the spectral weight distribution ($A_N$ vs. $N$) quantifies the e-ph coupling strength, and its spectral shift across $T_\mathrm{C}$ confirms the interplay between the polaronic effect and the magnetic order in bilayer \ce{CrI3}. Theoretically, the polaron system consisting of dispersionless LO phonons and charges is one of the few exactly solvable models in many-body physics \cite{RN62}, and the calculated polaron spectra can be well-described by a Frank-Condon model with a Poisson distribution function, $A_N = A_0\frac{e^{-\alpha}\alpha^N}{N!}$ \cite{RN63,RN64}, where $A_\mathrm{0}$ is the peak intensity of the original electronic band, $A_N$ is the peak intensity for the $N^\mathrm{th}$ replica band with $N$ phonon(s) dressed, and $\alpha$ is a constant related to the e-ph coupling in 3D (\textit{i.e.}, $\alpha_\mathrm{3D}$) that can be scaled by a factor of $3\pi/4$ for 2D (\textit{i.e.}, $\alpha_\mathrm{2D}$) \cite{RN65}. By fitting the extracted Lorentzian peak intensity profile at every temperature to a Poisson distribution function (see fits of 10 K and 70 K data in Fig. 3b), we achieve a comparable fitting quality to that for ARPES EDCs in polaron systems \cite{RN2, RN31} at every temperature and eventually arrive at the temperature dependence of $\alpha_\mathrm{2D}$ which remains nearly constant until the system is cooled to $T_\mathrm{C}$ and then increases by almost $50\%$ at the lowest available temperature 10 K of our setup (Fig. 3c). In addition to the anomalous enhancement of $\alpha_\mathrm{2D}$ across $T_\mathrm{C}$, the value of $\alpha_\mathrm{2D}$ = 1.5 at 10 K is the highest among known 2D polaron systems including graphene/BN heterostructures ($\alpha_\mathrm{2D}$ = 0.9) \cite{RN14} and bare \ce{SrTiO3} surfaces ($\alpha_\mathrm{2D}$ = 1.1) \cite{RN29}. 

It has been shown that bilayer \ce{CrI3} transitions from a layered AFM to FM with increasing out-of-plane magnetic field ($B_\bot$) above the critical value of $B_\mathrm{C}$ of 0.7 T \cite{RN6, RN8, RN20, RN21}. We then finally explore the evolution of polaronic effect across this magnetic phase transition by performing magnetic field-dependent Raman spectroscopy measurements. Here, we choose circularly polarized light to perform magnetic field-dependent measurements here in order to eliminate any Faraday effect from the optical components that are situated in close proximity to the strong magnetic field. Figure 4a shows Raman spectra taken at $B_\bot$ = 0 T and $\pm$1 T, below and above $B_\mathrm{C}$, respectively, in both RR and LL channels, where RR (LL) stands for the polarization channel selecting the right-hand (left-hand) circular polarization for both incident and scattered light (see Supplementary Section 6). At 0 T, the spectra are identical in RR and LL channels, consistent with zero net magnetization in the layered AFM state for bilayer \ce{CrI3} at $\mid B_\bot\mid<B_\mathrm{C}$. At $\pm$1 T, the spectra in RR and LL channels show opposite relative intensities under opposite magnetic field directions, due to the fact that the net magnetization in the FM state for bilayer \ce{CrI3} at $\mid B_\bot\mid>B_\mathrm{C}$ breaks the equivalence between the RR and LL channels. To better quantify the magnetic field dependence of the spectra, we measured Raman spectra in the RR and LL channels at $B_\bot$ from -1.4 T to 1.4 T every 0.1 T. We fit the spectrum at every magnetic field to extract $A_\mathrm{N}$ first, and then $A_\mathrm{0}$ and $\alpha_\mathrm{2D}$. Figure 4b shows that $A_\mathrm{0}$ has abrupt changes at $B_\bot = \pm$ 0.7 T in both RR and LL channels, consistent with the first order magnetic phase transition at $B_\mathrm{C}$. Furthermore, the magnetic field dependence of $A_\mathrm{0}$ shows an opposite trend in the RR channel from that in the LL channel, while the sum of $A_\mathrm{0}$ from both channels remain nearly constant to the varying magnetic field. This observation can be understood by that, under a time-reversal operation, the RR channel transforms into the LL channel while the direction of the net magnetization at $\mid B_\bot\mid>B_\mathrm{C}$ flips, resulting in that the Raman spectrum in the RR channel at $B_\bot >$ 0.7 T is equivalent to the spectrum in the LL channel at $B_\bot <$ -0.7 T. Figure 4c shows that $\alpha_\mathrm{2D}$ is magnetic-field independent, suggesting that the interlay magnetic order barely affect the e-ph coupling strength and that the in-plane long-range magnetic order is responsible for the strong enhancement of e-ph coupling at $T_\mathrm{C}$. This finding corroborates with the fact that the 120.6 $\mathrm{cm}^{-1}$ phonon is intralayer. 

Our further Raman spectroscopy studies on tri-layer, four-layer, and five-layer \ce{CrI3} show qualitatively same findings as those in bilayer \ce{CrI3} (see Supplementary Section 7) and again echoes with the in-plane nature of the 120.6 $\mathrm{cm}^{-1}$ $E_\mathrm{u}$ phonon and the intralayer charge transfer B exciton. Our data and analysis reveal the phonon-dressed electronic states and suggest the polaronic character of excitons in 2D \ce{CrI3} which arises from the strong coupling between the lattice and charge degrees of freedom and is dramatically modified by the spin degree of freedom of \ce{CrI3}. The exceptionally high number of phonon-dressed electronic state replicas (up to $N$ = 8) further suggests 2D \ce{CrI3} as an outstanding platform to explore nontrivial phases out of phonon-Floquet engineering, while the significant coupling to the spin degree of freedom adds an extra flavor whose impact on the phonon-Floquet states has not been studied. For example, one can imagine creating topological states through the band inversion between the phonon-dressed replicas of \ce{CrI3} and the electronic state of a material in close proximity. \\

\noindent \textbf{Methods}

\noindent \textbf{Sample fabrication.} \ce{CrI3} single crystals were grown by the chemical vapor transport method, as detailed in Ref. \cite{RN40}. Bilayer \ce{CrI3} samples were exfoliated in a nitrogen-filled glovebox. Using a polymer-stamping transfer technique inside the glove box, bilayer \ce{CrI3} flakes were sandwiched between two few-layer hBN flakes and transferred onto \ce{SiO2}/Si substrates and sapphire substrates for Raman spectroscopy and PL/linear absorption spectroscopy measurements, respectively. \\

\noindent \textbf{Linear absorption spectroscopy.} A bilayer \ce{CrI3} sample on a sapphire substrate was mounted in a closed-cycle cryostat for the temperature-dependent absorption spectroscopy measurements. A broadband tungsten lamp was focused onto the sample via a $50\times$ long working distance objective. The transmitted light was collected by another objective and coupled to a spectrometer with a spectral resolution of 0.2 nm. The absorption spectra were determined by $1-\frac{I_\mathrm{sample} (\lambda)}{I_\mathrm{substrate} (\lambda)}$, where $I_\mathrm{sample} (\lambda)$ and $I_\mathrm{substrate} (\lambda)$ were the transmitted intensity through the combination of sample and substrate and through the bare substrate, respectively.\\

\noindent \textbf{Photoluminescence spectroscopy.} PL spectra were acquired from the same bilayer \ce{CrI3} sample where we carried out linear absorption measurements. The sample was excited by a linearly polarized 633 nm laser focused to a $\sim\SI{2}{\micro\meter}$ spot. A power of $\SI{30}{\micro\watt}$ was used, which corresponds to a similar fluence reported in the literature ($\sim\SI{10}{\micro\watt}$ over a $\sim\SI{1}{\micro\meter}$-diameter spot) \cite{RN3}. Transmitted right-handed circularly polarized PL signal was dispersed by a 600 groves/mm, 750 nm blaze grating, and detected by an InGaAs camera.\\ 

\noindent \textbf{Raman spectroscopy.} Resonant micro-Raman spectroscopy measurements were carried out using a 633 nm excitation laser for the data in the main text and 473 nm, 532 nm, and 785 nm excitation laser for data in Supplementary Section 5. The incident beam was focused by a $40\times$ objective down to $\sim\SI{3}{\micro\meter}$ in diameter at the sample site, and the power was kept at $\sim\SI{80}{\micro\watt}$. The scattered light was collected by the objective in a backscattering geometry, then dispersed by a Horiba Labram HR Raman Evolution Raman spectrometer, and finally detected by a thermoelectric cooled CCD camera. A closed-cycle helium cryostat is interfaced with the micro-Raman system for the temperature-dependent measurements. All thermal cycles were performed at a base pressure lower than $7 \times 10^{-7}$ mbar. In addition, a cryogen-free magnet is integrated with the low temperature cryostat for the magnetic field-dependent measurements. In this experiment, the magnetic field was applied along the out-of-plane direction and covered a range of -1.4 to 1.4 Tesla. In order to avoid the Faraday rotation of linearly polarized light as it transmits through the objective under the magnetic field, we used circularly polarized light to perform the magnetic field-dependent Raman measurements. \\

\noindent \textbf{Fitting polaron Raman spectroscopy.} For every sample, we have taken temperature and magnetic field dependent Raman spectra on the hBN/\ce{SiO2}/Si substrate with the same experimental conditions as that on the \ce{CrI3} flakes. The Raman spectra from the substrate, an extremely gradual background with a Si phonon peak at $\sim$525 $\mathrm{cm}^{-1}$, shows no dependence on temperature (over the range of 10--70 K) and magnetic field (0--2.2 T). To fit the polaron Raman spectra of \ce{CrI3} flakes, we follow the routine described below. (I) we fit the Si phonon peak at $\sim$525 $\mathrm{cm}^{-1}$ in both spectra taken on the \ce{CrI3} thin flake and the bare substrate to extract the Si peak intensity, $I_\mathrm{Si}^\mathrm{sample}$ and $I_\mathrm{Si}^\mathrm{substrate}$. (II) we multiply the background spectrum by a factor of $I_\mathrm{Si}^\mathrm{sample}/I_\mathrm{Si}^\mathrm{substrate}$, which is nearly 1, and then subtract off the factored background from the raw Raman spectrum of sample. (III) we fit the sharp \ce{CrI3} phonon peaks with Lorentzian functions and substrate subtract their fitted functions from the background free Raman spectrum from step II. This leads to a clean polaron spectrum with only periodic broad modes. (IV) we fit the clean polaron spectrum from step III with a sum of multiple Lorentzian functions, $\sum_N\frac{A_N(\varGamma_N/2)^2}{(\omega-{\omega}_N)^2+(\varGamma_N/2)^2}+C$. For the neatness of the data presentation in Fig. 2 and 3, we only show the fitted line from step IV in the plots. \\

\newpage
\noindent\textbf{Acknowledgements} \\
\noindent We thank X. Xu, M. Kira, R. Merlin, X. Qian, and H. Wang for useful discussions. L. Zhao acknowledges support by NSF CAREER Grant No. DMR-1749774. R. He acknowledges support by NSF CAREER Grant No. DMR-1760668 and NSF MRI Grant No. DMR-1337207. K. Sun acknowledges support through NSF Grant No. NSF-EFMA-1741618. A. W. Tsen acknowledges support by NSERC Discovery grant RGPIN-2017-03815 and the Korea-Canada Cooperation Program through the National Research Foundation of Korea (NRF) funded by the Ministry of Science, ICT and Future Planning (NRF-2017K1A3A1A12073407). This research was undertaken, thanks in part to funding from the Canada First Research Excellence Fund. H. Lei acknowledges support by the National Key R\&D Program of China (Grant No. 2016YFA0300504), the National Natural Science Foundation of China (No. 11574394, 11774423, and 11822412), the Fundamental Research Funds for the Central Universities, and the Research Funds of Renmin University of China (15XNLQ07, 18XNLG14, and 19XNLG17). H. Deng and J. Horng acknowledge support by the Army Research Office under Awards W911NF-17-1-0312. \\

\noindent\textbf{Author contributions}\\
\noindent W. J., R. H., and L. Z. conceived this project; S. T., Y. F., and H. L. synthesized and characterized the bulk \ce{CrI3} single crystals; H. H. K., B. Y., F. Y., and A. W. T. fabricated and characterized the few-layer samples; Z. Y., G. Y., L. R. and R. H. performed the Raman measurements; J. H., W. J., and H. D. performed the linear absorption and photoluminescence spectroscopy measurements; W. J., X. L., R. H., and L. Z. analyzed the data with discussions with K. S.; W. J., X. L.,  R. H., and L. Z. wrote the paper and all authors participated in the discussions of the results.  \\

\noindent\textbf{Data availability}\\
\noindent The datasets generated and/or analyzed during the current study are available from the corresponding author on reasonable request.\\

\noindent\textbf{Competing interests}\\
\noindent The authors declare no competing interests.\\


\bibliographystyle{apsrev4-1}
\nocite{apsrev41Control}
\bibliography{arXiv.bib} 

\newpage
\begin{figure}
\includegraphics[scale=0.8]{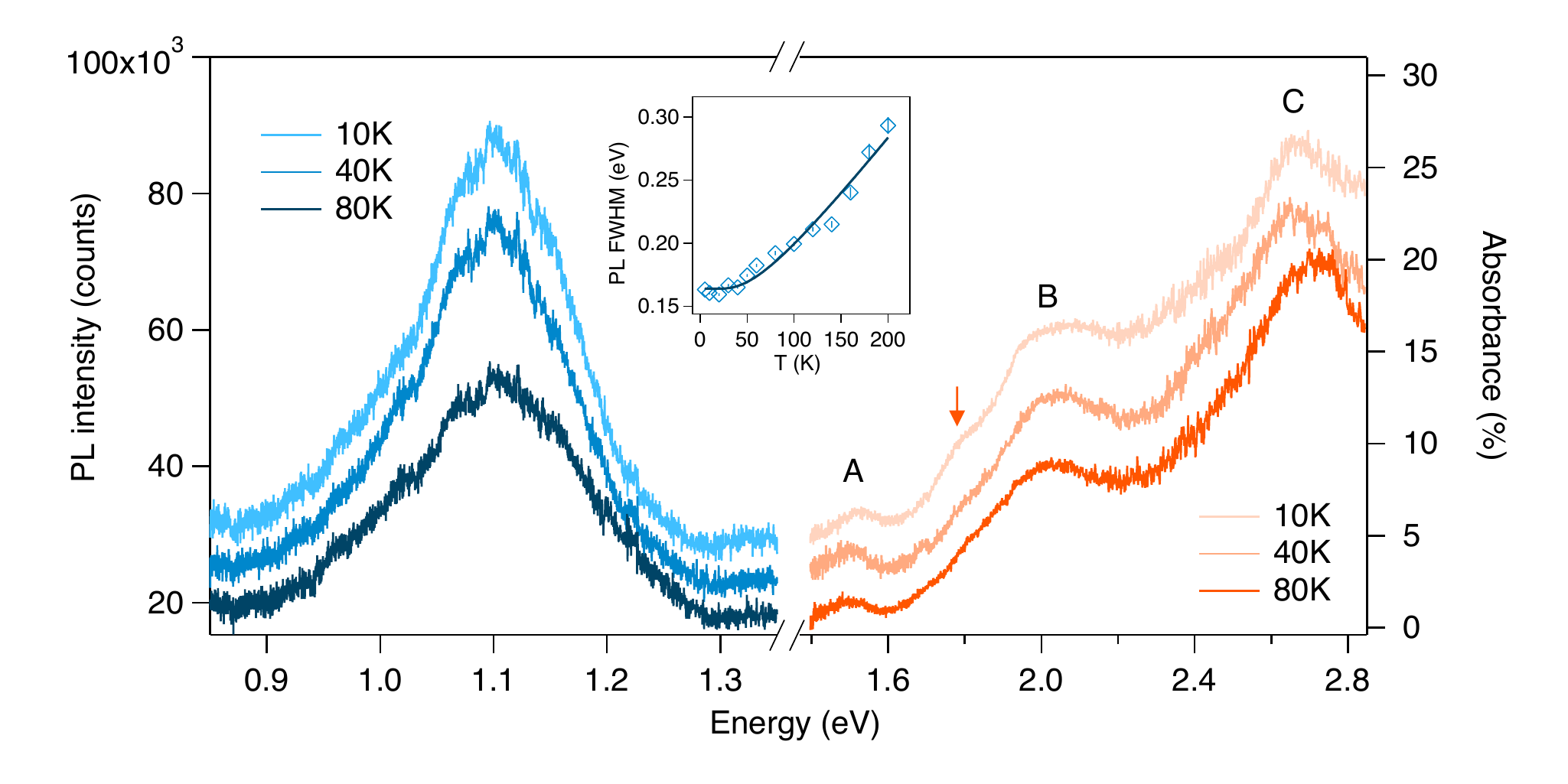}
\end{figure}
\begin{footnotesize}
\noindent \textbf{Fig. 1. Exciton transitions in bilayer \ce{CrI3}.} Photoluminescence (PL) (left, blue) and absorption (right, orange) spectra of a bilayer \ce{CrI3} encapsulated between two hBN flakes and placed on the sapphire substrate, at 10, 40, and 80 K. A, B, and C denote three main exciton transitions at 1.51, 1.96, and 2.68 eV, and the orange arrow mark a shoulder mode at 1.79 eV appearing only at low temperature. Spectra at 10 K and 40 K are offset vertically for clarity. Inset shows the fitted full width at half maximum (FWHM) of PL spectra as a function of temperature, $\varGamma(T) = \varGamma_0 + \frac{\gamma}{\mathrm{exp}(\frac{\hbar\omega_\mathrm{LO}}{{k_B}T})-1}$, (diamond symbols) and its fitting to the functional form with the first and second terms for homogeneous and e-ph coupling-induced broadening, respectively. Error bars indicate one standard error in fitting the FWHM of PL spectra.
\end{footnotesize}

\newpage
\begin{figure}
\includegraphics[scale=0.7]{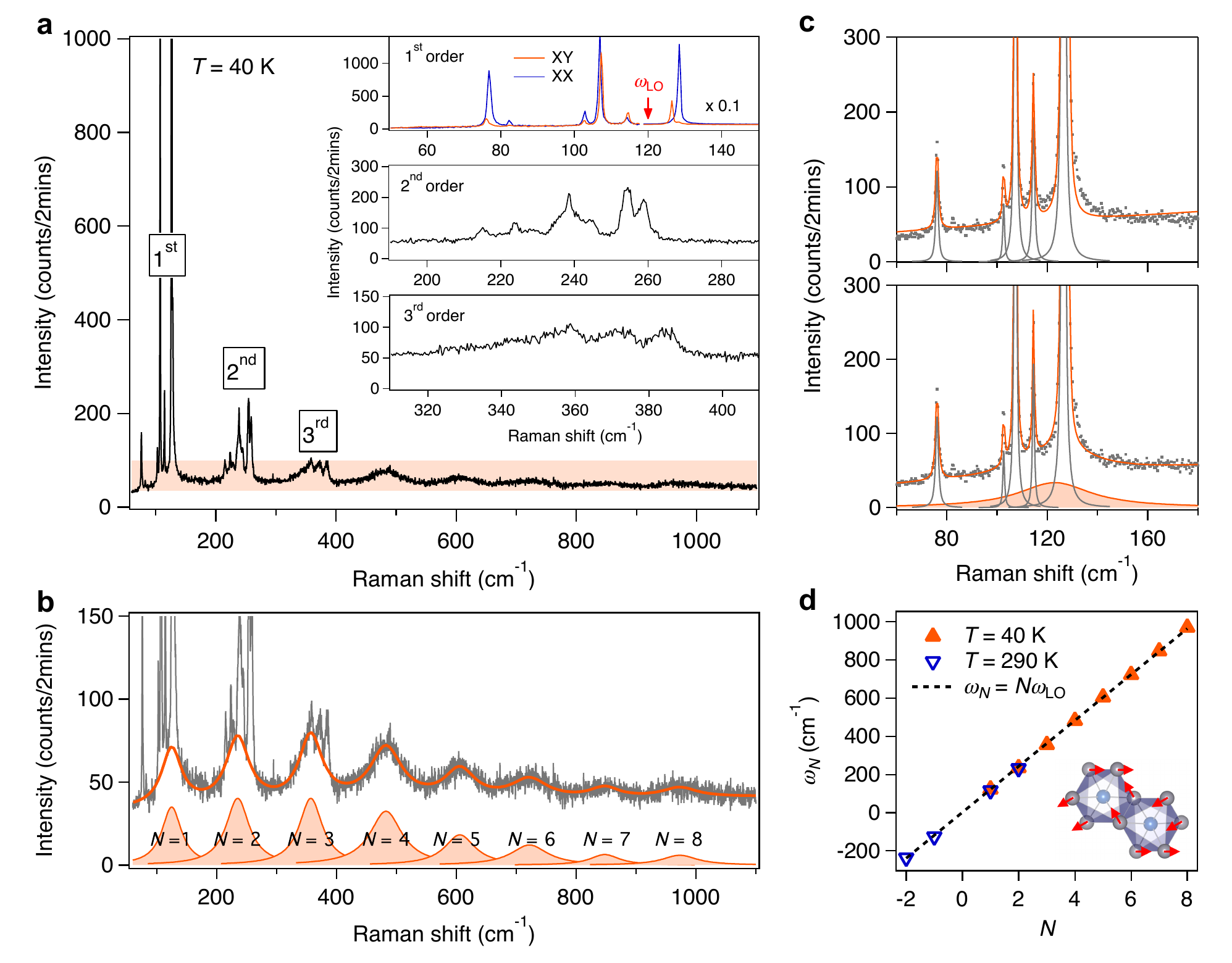}
\end{figure}
\begin{footnotesize}
\noindent \textbf{Fig. 2. Polaronic character of exciton dynamics in bilayer \ce{CrI3}.} \textbf{a}. Raman spectrum of bilayer \ce{CrI3} acquired in the linearly crossed polarization channel at 40 K using a 633 nm laser. Inset shows $1^\mathrm{st}$ order single phonon modes (linearly parallel/crossed, \textit{i.e.}, XX/XY, channel in blue/red), $2^\mathrm{nd}$ order two-phonon modes, and $3^\mathrm{rd}$ order three-phonon modes. The red arrow indicates the Raman-inactive longitudinal optical (LO) phonon frequency of importance, $\omega_\mathrm{LO}$. The spectral intensities in the frequency range above 118 $\mathrm{cm}^{-1}$ are scaled by a factor of 0.1. \textbf{b}. Zoom-in of the orange shaded area in the spectrum in \textbf{a}. The solid orange line overlaid on the raw data is the fit to the periodic pattern in the Raman spectrum with a sum of eight Lorentzian profiles indexed from $N$ = 1 to $N$ = 8 and a constant background, \textit{i.e.}, $\sum_N\frac{A_N(\varGamma_N/2)^2}{(\omega-{\omega}_N)^2+(\varGamma_N/2)^2}+C$. \textbf{c}. Fit of the Raman spectrum over the $1^\mathrm{st}$ order spectral range without (upper panel) and with (lower panel) the consideration of the $N$ = 1 broad mode (shaded Lorentzian profile in the lower panel). \textbf{d}. Plot of the fitted central frequency ($\omega_N$) of the $N^\mathrm{th}$ Lorentzian profile in data taken at 40 K (solid orange triangle) and 290 K (hollow blue triangle). Dashed line is a linear fit ($\omega_N=N\omega_\mathrm{LO}$) to the plot that yields a slope of $\omega_\mathrm{LO}=120.6 \pm 0.9 \mathrm{cm}^{-1}$. Inset shows the atomic displacement of the LO phonon mode. 
\end{footnotesize}

\newpage
\begin{figure}
\includegraphics[scale=0.7]{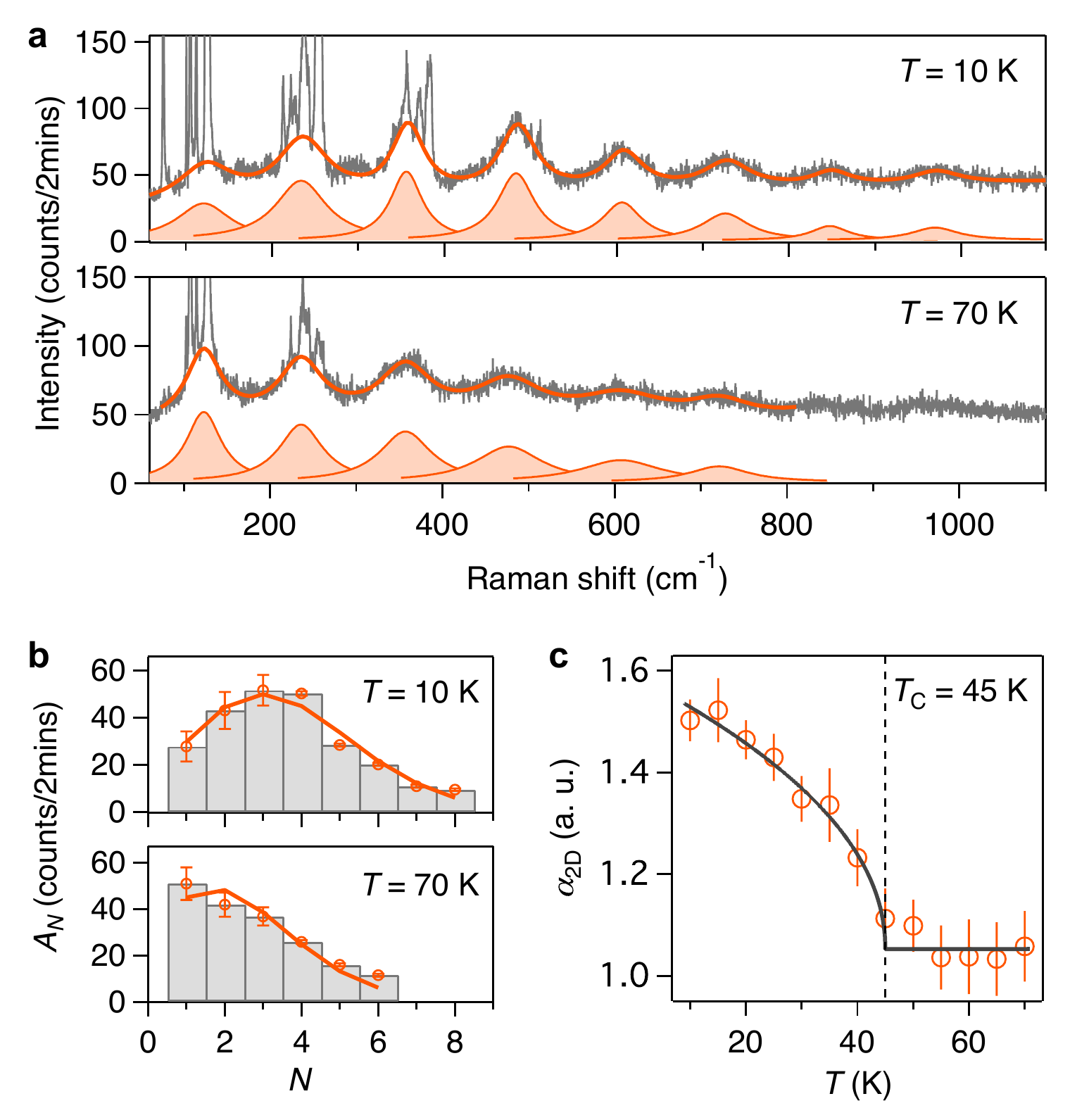}
\end{figure}
\begin{footnotesize}
\noindent \textbf{Fig. 3. Enhanced electron-phonon coupling across the magnetic onset $T_\mathrm{C}$ in bilayer \ce{CrI3}.} \textbf{a}. Raman spectra of bilayer \ce{CrI3} acquired at 10 K and 70 K, respectively. Solid orange lines are fits to the raw Raman spectra, using a sum of $N$ Lorentzian profiles and a constant background, \textit{i.e.}, $\sum_N\frac{A_N(\varGamma_N/2)^2}{(\omega-{\omega}_N)^2+(\varGamma_N/2)^2}+C$. \textbf{b}. Histogram plots of the fitted Lorentzian peak intensity ($A_N$) as a function of order index ($N$) at 10 K and 70 K. Solid curves are fits of the peak intensity profiles to the Poisson distribution functions, $A_N = A_0\frac{e^{-\alpha}\alpha^N}{N!}$. Error bars represent one standard deviation that convolutes the standard errors from the fittings to the local individual and global multiple Lorentzian profiles. \textbf{c}. Plot of two-dimensional electron-phonon coupling constant ($\alpha_\mathrm{2D}$) as a function of temperature. The dashed vertical line marks the magnetic onset $T_\mathrm{C}$ = 45 K, and the solid line is the fit to the functional form $\alpha_\mathrm{2D}(T)$ = $\begin{cases} A\sqrt{T_\mathrm{C}-T}+B; & T\leq T_\mathrm{C} \\ B; &T>T_\mathrm{C} \end{cases}$. Error bars in \textbf{c} represent one standard error in the Poisson fitting.
\end{footnotesize}

\newpage
\begin{figure}
\includegraphics[scale=0.8]{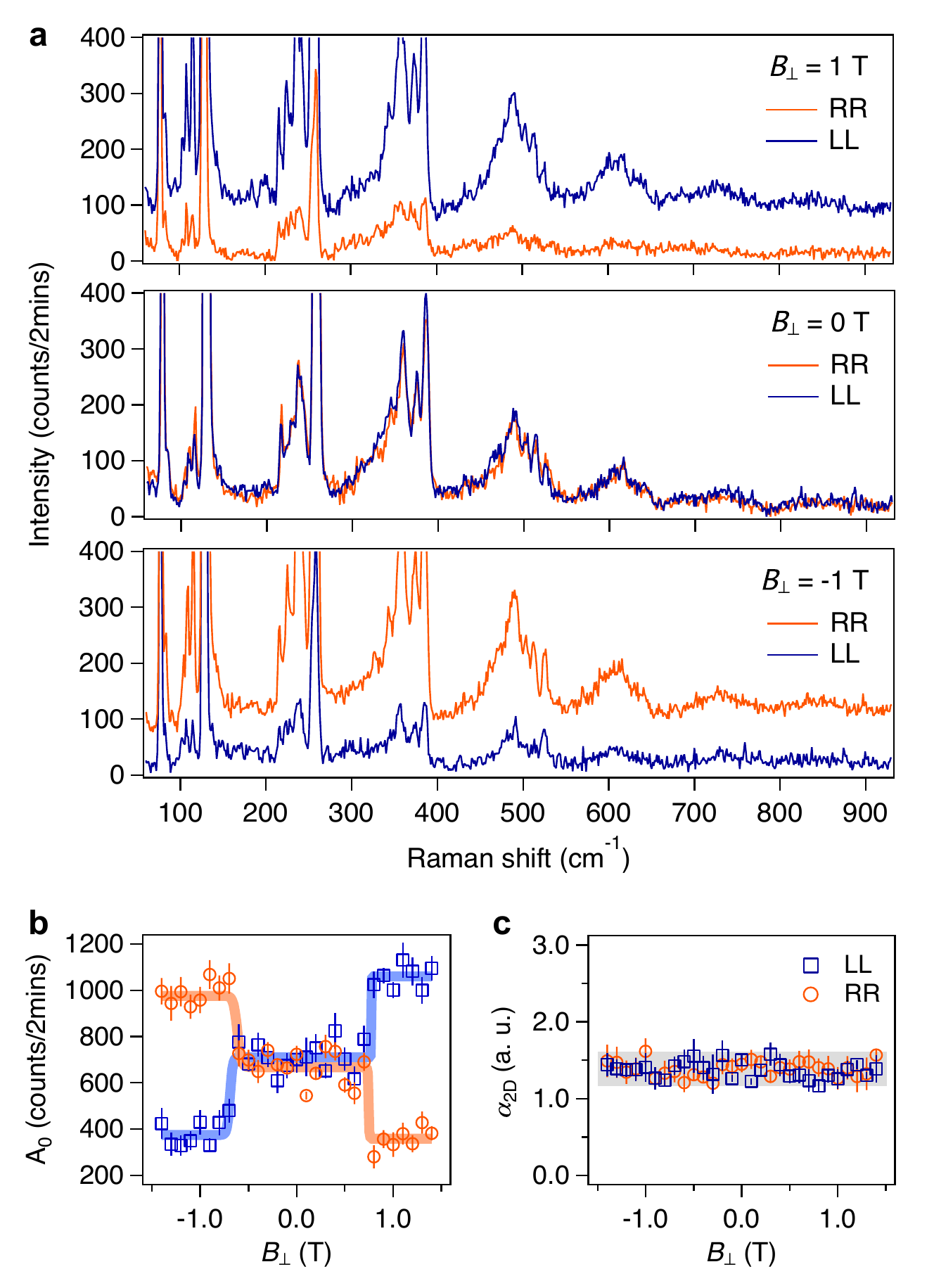}
\end{figure}
\begin{footnotesize}
\noindent \textbf{Fig. 4. Evolution of polaronic effect across the magnetic phase transition from the layered AFM to FM in bilayer \ce{CrI3}.} \textbf{a}. Raman spectra of bilayer \ce{CrI3} acquired at 10 K in the circularly parallel polarization channels, RR and LL, where RR (LL) stands for that both incident and scattered polarizations are selected to be right (left) circularly polarized, with an applied out-of-plane magnetic field ($B_\bot$) of 1 T (top), 0 T (middle), and -1 T (bottom), respectively. \textbf{b}--\textbf{c}. Plots of the Poisson fit amplitude $A_\mathrm{0}$ (\textbf{b}) and electron-phonon coupling strength $\alpha_\mathrm{2D}$ (\textbf{c}) as a function of the applied $B_\bot$ in RR (orange data points) and LL (blue) channels. Solid lines are step (orange and blue in \textbf{b}) and linear (gray in \textbf{c}) function fits to the magnetic field dependence of $A_\mathrm{0}$ and $\alpha_\mathrm{2D}$, respectively. Error bars are defined as one standard error of the fitting parameters.
\end{footnotesize}

\newpage
\noindent \textit{Supplementary Information}\\
\begin{center}
\textbf{\large Observation of the polaronic character of excitons in a two-dimensional semiconducting magnet \ce{CrI3}}\\

Wencan Jin$,^{1, \ast}$ Hyun Ho Kim$,^{2, \dagger}$ Zhipeng Ye$,^3$ Gaihua Ye$,^3$ Laura Rojas$,^3$ Xiangpeng Luo$,^1$ Bowen Yang$,^2$ Fangzhou Yin$,^2$ Jason Shih An Horng$,^1$  Shangjie Tian$,^4$ Yang Fu$,^4$ Hui Deng$,^1$ Hechang Lei$,^4$ Adam W. Tsen$,^2$ Kai Sun$,^1$ Rui He$^{3, \ddagger}$ and Liuyan Zhao$,^{1, \S}$\\

\textit{$^1$Department of Physics, University of Michigan, 450 Church Street,\\ Ann Arbor, Michigan 48109, USA}\\

\textit{$^2$Institute for Quantum Computing, Department of Chemistry, \\and Department of Physics and Astronomy, University of Waterloo,\\ Waterloo, 200 University Ave W, Ontario N2L 3G1, Canada}

\textit{$^3$Department of Electrical and Computer Engineering, 910 Boston Avenue, \\Texas Tech University, Lubbock, Texas 79409, USA}

\textit{$^4$Department of Physics and Beijing Key Laboratory of \\Opto-electronic Functional Materials \& Micro-nano Devices, \\Renmin University of China, Beijing 100872 China}
\end{center}

\noindent \textbf{Table of Contents}\\
\noindent{S1. First-order phonon modes in bilayer \ce{CrI3}}\\
\noindent{S2. Anti-Stokes and Stokes polaron Raman spectra taken at 290 K in bilayer \ce{CrI3}}\\
\noindent{S3. The presence of broad mode rigorously confirmed by the bootstrap method}\\
\noindent{S4. Periodic pattern of broad modes in Raman spectrum observed in bulk \ce{CrI3}}\\
\noindent{S5. Wavelength dependent Raman spectra of bilayer \ce{CrI3}}\\
\noindent{S6. Polaronic effect in Raman spectra of bilayer \ce{CrI3} in all four circular polarization channels}\\
\noindent{S7. Temperature and magnetic field dependent Raman measurements for tri-layer, four-layer, and five-layer \ce{CrI3}}\\

\noindent{$\ast$ current affiliation: Department of Physics, Auburn University, 380 Duncan Drive, Auburn, AL 36849, USA}\\
\noindent{$\dagger$ current affiliation: School of Materials Science and Engineering, Kumoh National Institute of Technology, Gumi,
Gyeongbuk 39177, Korea}\\
\noindent{$\ddagger$ rui.he@ttu.edu, $\S$ lyzhao@umich.edu}


\newpage
\noindent \textbf{S1. First-order phonon modes in bilayer \ce{CrI3}}\\

\noindent Figure S1 shows the Raman spectra in the frequency range of 60--150 $\mathrm{cm}^{-1}$. The first- order phonon modes are assigned to be either $A_\mathrm{g}$ ($A_1$ -- $A_3$) or $E_\mathrm{g}$ ($E_1$ -- $E_4$) symmetries under the $C_\mathrm{3i}$ point group. The two modes that appear in the magnetic phase are labeled as $M_1$ and $M_2$. This assignment is consistent with earlier works.\\

\begin{figure}[!h]
\includegraphics[scale=0.8]{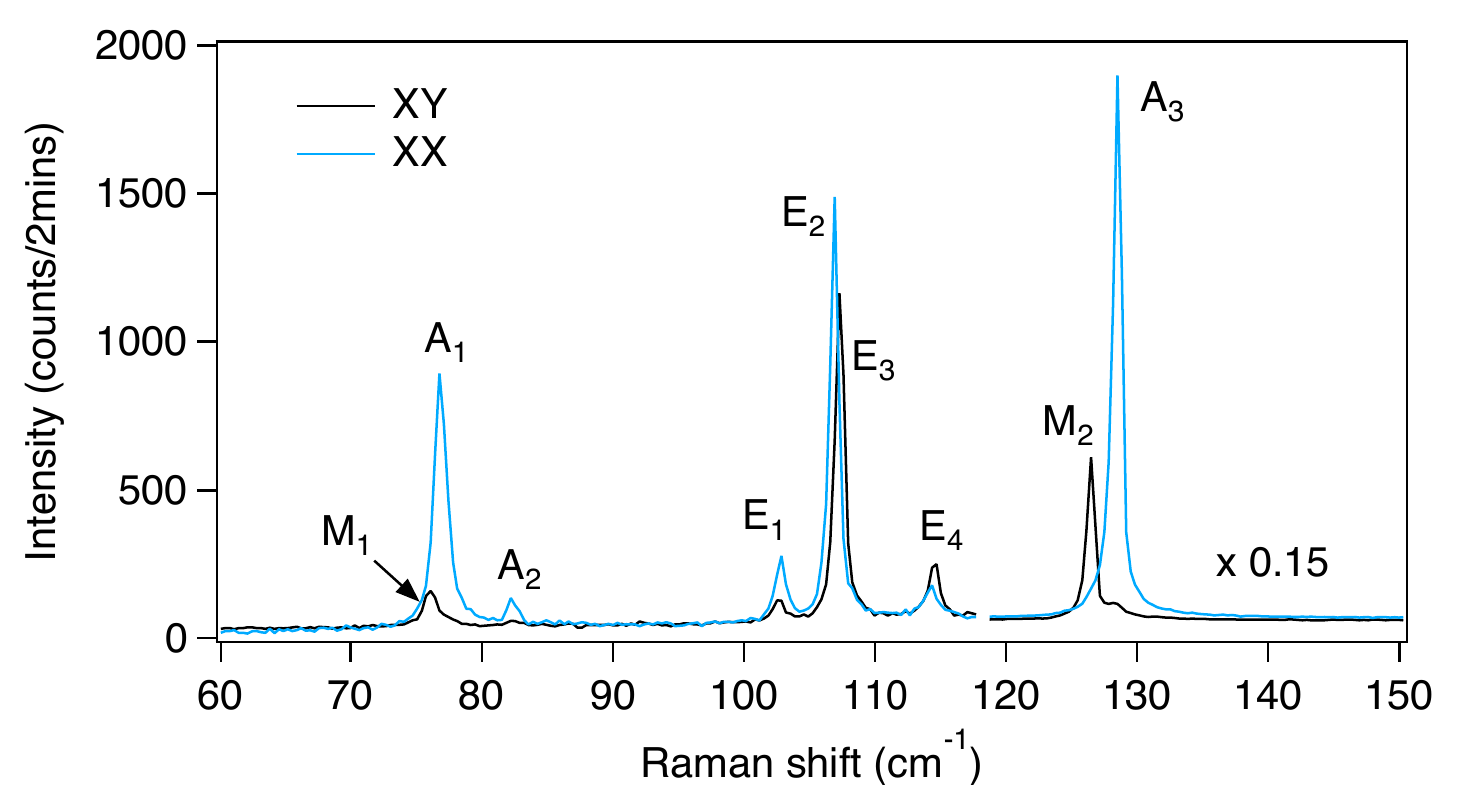}
\end{figure}
\begin{footnotesize}
\noindent \textbf{Fig. S1.} Raman spectra of bilayer \ce{CrI3} in the range of 60--150 $\mathrm{cm}^{-1}$ acquired in parallel (XX) and crossed (XY) linear polarization channels at 40 K using a 633 nm laser. The spectra in the frequency range above 118 $\mathrm{cm}^{-1}$ are scaled by a factor of 0.15. The first-order phonon modes are labeled as $A_1$ -- $A_3$ and $E_1$ -- $E_4$, of $A_\mathrm{g}$ and $E_\mathrm{g}$ symmetry under the $C_\mathrm{3i}$ point group.  \\
\end{footnotesize}

\newpage
\noindent \textbf{S2. Anti-Stokes and Stokes polaron Raman spectra taken at 290 K in bilayer \ce{CrI3}}\\

\noindent Figure S2 shows the Raman spectrum acquired in the frequency range from -1000 to 1000 $\mathrm{cm}^{-1}$ at 290 K. The periodic pattern is on the anti-Stokes side up to -$2^\mathrm{nd}$ order. The broad mode centered at 0 $\mathrm{cm}^{-1}$ is attributed to the quasi-elastic scattering (QES) due to spin fluctuations. The broad and skewed background was subtracted before fitting the periodic pattern, and is absent over the temperature range of most interest and relevance in the main text (10--70 K).\\

\begin{figure}[!h]
\includegraphics[scale=1]{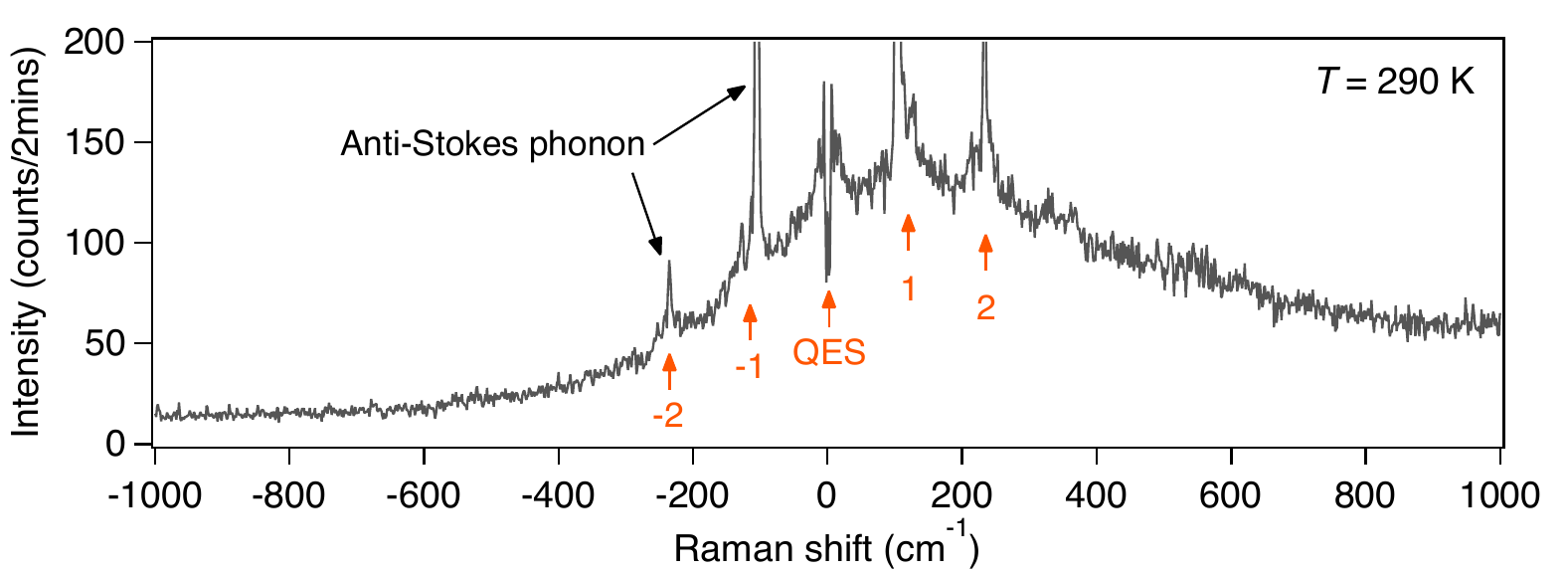}
\end{figure}
\begin{footnotesize}
\noindent \textbf{Fig. S2.} Raman spectrum extending to anti-Stokes side taken on a bilayer \ce{CrI3} at $T$ = 290 K in linearly crossed polarization channel. Arrows with the numbers denote the indexes of polaron modes. QES stands for quasi-elastic scattering. \\
\end{footnotesize}

\newpage
\noindent \textbf{S3. The presence of broad mode rigorously confirmed by the bootstrap method}\\

\noindent It is nontrivial to identify the broad modes in the polaron spectra over the spectral ranges where the strong, sharp phonon modes are present, in particular, the spectral range of 60--160 $\mathrm{cm}^{-1}$ covering the $1^\mathrm{st}$ order phonon modes and $N$ = 1 broad mode.\\

\noindent Bootstrap method is a standard statistical approach to testify an assumption using computational calculations. Here, the to-be-tested assumption is that there is no $N$ = 1 broad mode over the spectral range of 60--160 $\mathrm{cm}^{-1}$. By performing the calculations described below, we confidently reject this assumption, and state that it is statistically significant to have $N$ = 1 broad mode present over 60--160 $\mathrm{cm}^{-1}$ in the Raman spectra of bilayer \ce{CrI3}.\\

\noindent In particular, we take the following steps to carry out the bootstrap-based test on a representative Raman spectrum taken at 40 K (Fig. 2 in the main text).\\

\noindent Step 1: Fit the raw data using two methods: Method 1: five sharp Lorentzian profiles to account for five sharp phonon modes; and Method 2: five sharp Lorentzian profiles for phonons and one broad Lorentzian profile for $N$ = 1 polaron mode.\\

\noindent From the fitting result in Step 1 panel of Fig. S3 (reproduced from Fig. 2c of the main text), Method 2 is visibly better than Method 1. Quantitatively, the sum square error (SSE) and root mean square error (RMSE) obtained in Method 2 is smaller (better) than those of Method 1 by $\Delta\mathrm{SSE}_0$ = 2.1$\times10^4$ and $\Delta\mathrm{RMSE}_0$ = 1.32, respectively. These numbers immediately lead to a question whether the improvement from Method 1 to Method 2 is statistically significant. It can only be addressed by comparing these two numbers with their counterparts in a situation where a $N$ = 1 broad Lorentzian mode is indeed absent.\\

\noindent Step 2: Simulate the data to mimic the situation where the $N$ = 1 broad mode is absent, by generating the spectra with the fitted curve in Method 1 plus the random Gaussian noise of standard error same as the RMSE in Method 1. One of the simulated spectra is shown in Fig. S3 (Step 2 panel).\\

\noindent Step 3: Fit the simulated data from Step 2 with both Method 1 and Method 2, compute the SSE difference ($\Delta$SSE) and RMSE difference ($\Delta$RMSE) between the two methods, and repeat this for 100 simulated spectra in Step 2. By comparing $\Delta\mathrm{SSE}_0$ = 2.1$\times10^4$ and $\Delta\mathrm{RMSE}_0$ = 1.32 to the distribution of $\Delta$SSE and $\Delta$RMSE (Step 3 panel in Fig. S3), respectively, both $\Delta\mathrm{SSE}_0$ and $\Delta\mathrm{RMSE}_0$ are clearly greater than the mean of $\Delta$SSE and $\Delta$RMSE by more than twice of their corresponding standard deviations. This confirms that the reduction of SSE and RMSE in fitting the raw data with Method 2 (by adding an extra broad Lorentzian mode as compared to Method 1) is statistically significant, and therefore the presence of the $N$ = 1 broad mode is significant.\\

\begin{figure}[!h]
\includegraphics[scale=1]{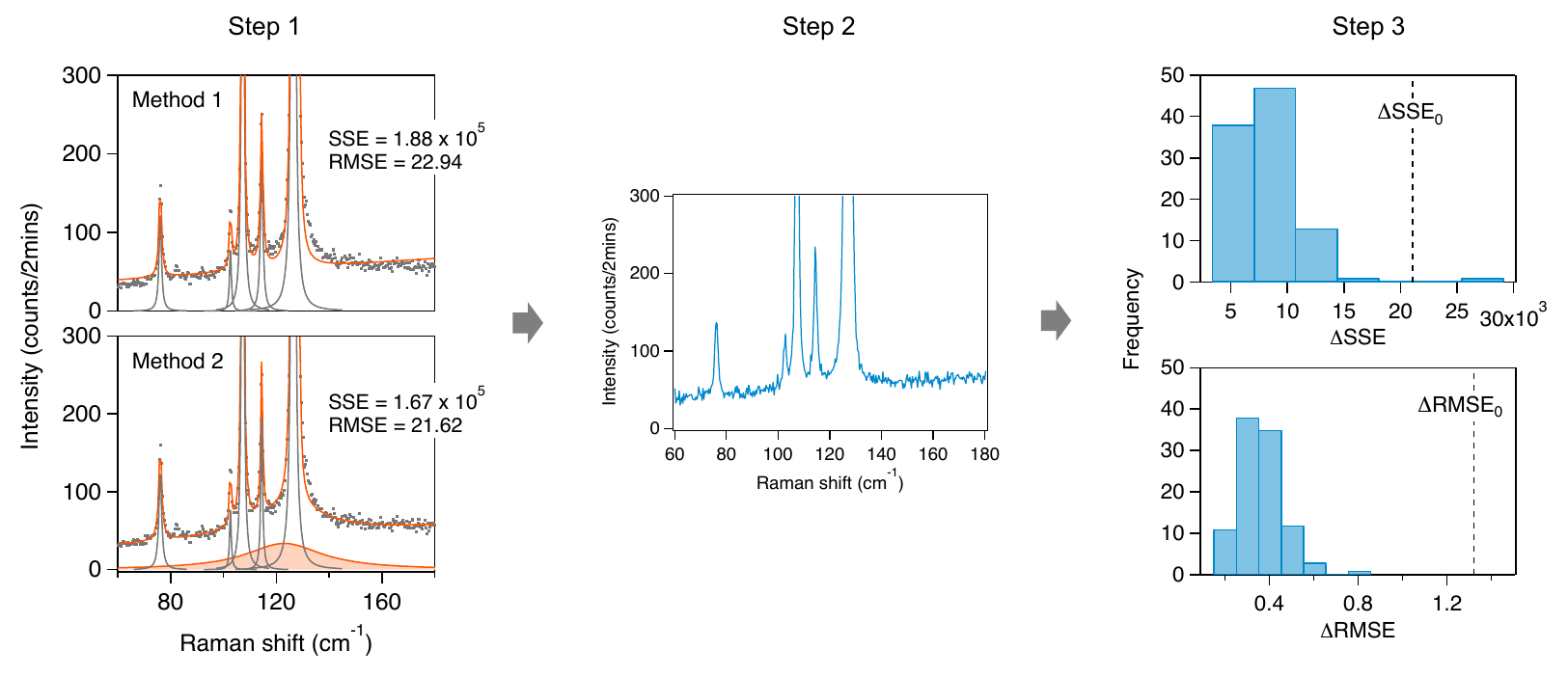}
\end{figure}
\begin{footnotesize}
\noindent \textbf{Fig. S3.} Step 1: Fit Raman spectra in the frequency range of 60--180 $\mathrm{cm}^{-1}$ using Method 1 (five sharp Lorentzian profiles, upper panel) and Method 2 (five sharp Lorentzian profiles plus a broad mode, lower panel), respectively; Step 2: Simulate the spectrum for the situation where there is no broad mode present; Step 3: Fit the reconstructed spectrum from Step 2 using Method 1 and Method 2, and plot the histogram distribution of the $\Delta$SSE (upper panel) and $\Delta$RMSE  (lower panel) for 100 simulated spectra. Dashed lines mark the values of $\Delta\mathrm{SSE}_0$ and $\Delta\mathrm{RMSE}_0$ exacted from the raw data fitting in Step 1.  \\
\end{footnotesize}

\newpage
\noindent \textbf{S4. Periodic pattern of broad modes in Raman spectrum observed in bulk \ce{CrI3}}\\

\noindent We performed Raman spectroscopy measurements on a freshly cleaved \ce{CrI3} bulk crystal using identical experimental conditions as that taken on bilayer \ce{CrI3} samples. Figure S4 shows a Raman spectrum acquired at 70 K. A similar periodic pattern as that in bilayer \ce{CrI3} is observed in bulk \ce{CrI3} and is fitted by a summation of Lorentzian profiles of the form $\sum_N\frac{A_N(\varGamma_N/2)^2}{(\omega-{\omega}_N)^2+(\varGamma_N/2)^2}+C$. Inset shows the plot of the fitted central frequency $\omega_N$ as a function of $N$, whose linear fit provides the periodicity of 121.3 $\pm$ 1.2 $\mathrm{cm}^{-1}$. This periodicity, \textit{i.e.}, the frequency of the participating polar LO phonons in the polaronic effect, is slightly greater than that of 120.6 $\pm$ 0.9 $\mathrm{cm}^{-1}$ in bilayer \ce{CrI3}. This can be attributed to the additional interlayer coupling on both sides of a layer in bulk \ce{CrI3}, which is supported by similar amount of blue shift for the Raman active phonon modes. The consistency between bulk and bilayer \ce{CrI3} definitively rules out the possibility of the polaron in bilayer \ce{CrI3} resulting from the interfacial coupling with the hBN capsulation layers or the \ce{SiO2}/Si substrate. \\

\begin{figure}[!h]
\includegraphics[scale=0.7]{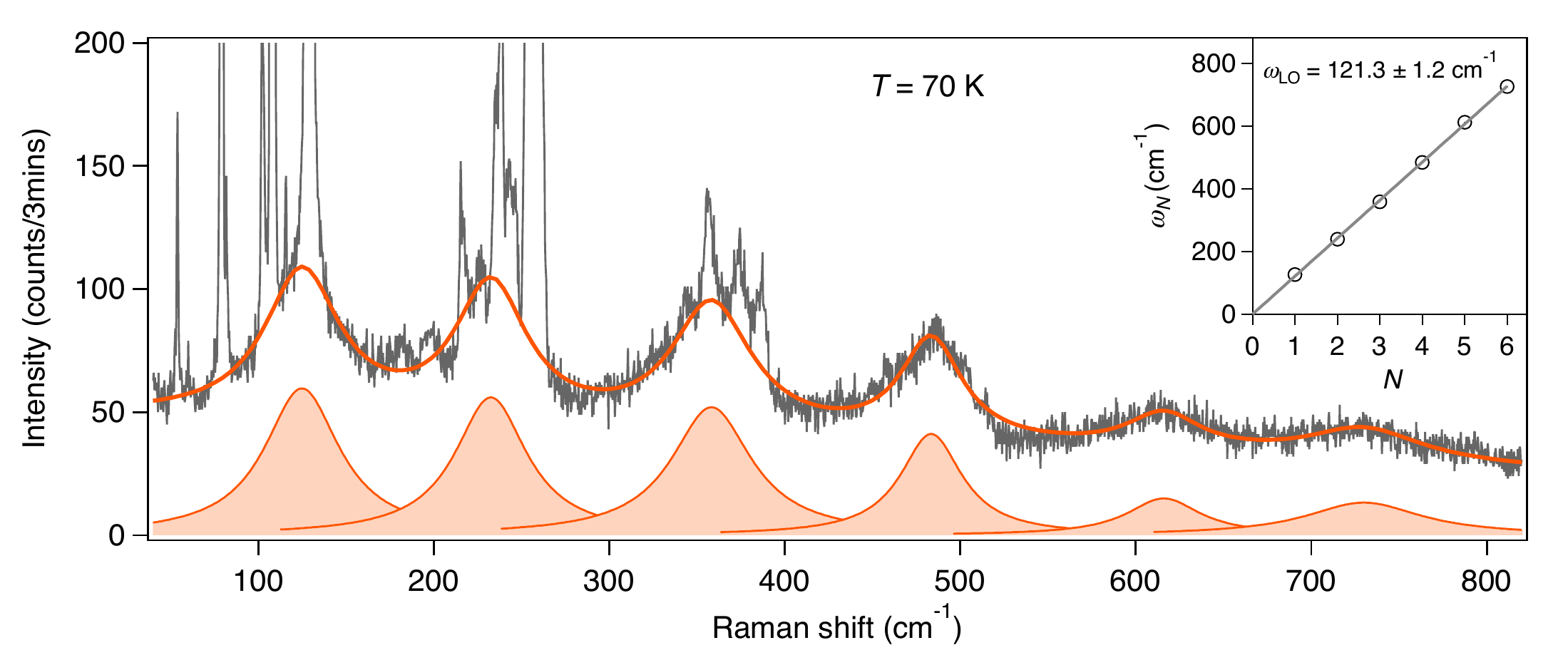}
\end{figure}
\begin{footnotesize}
\noindent \textbf{Fig. S4.} Raman spectrum of bulk \ce{CrI3} acquired in the crossed linear polarization channel at 70 K using a 633 nm laser. Lorentzian profiles (orange shade) are fits to the multiples. Inset shows the plot of the central frequencies ($\omega_N$) as a function of the index $N$. The frequency of the LO phonon ($\omega_\mathrm{LO}$) is extracted by a linear fit to $\omega_N$ (solid line). 
\end{footnotesize}

\newpage
\noindent \textbf{S5. Wavelength dependent Raman spectra of bilayer \ce{CrI3}}\\

\noindent We carried out Raman spectroscopy measurements on bilayer \ce{CrI3} using 785 nm, 532 nm, and 473 nm laser, whose energy locations are marked in Fig. S5a. Polaronic character is only observed at 633 nm (B exciton) but not at 785 nm and 473 nm excitations (see Fig. S5b) due to two reasons, a scientific one and a technical one.

\begin{figure}[!h]
\includegraphics[scale=0.8]{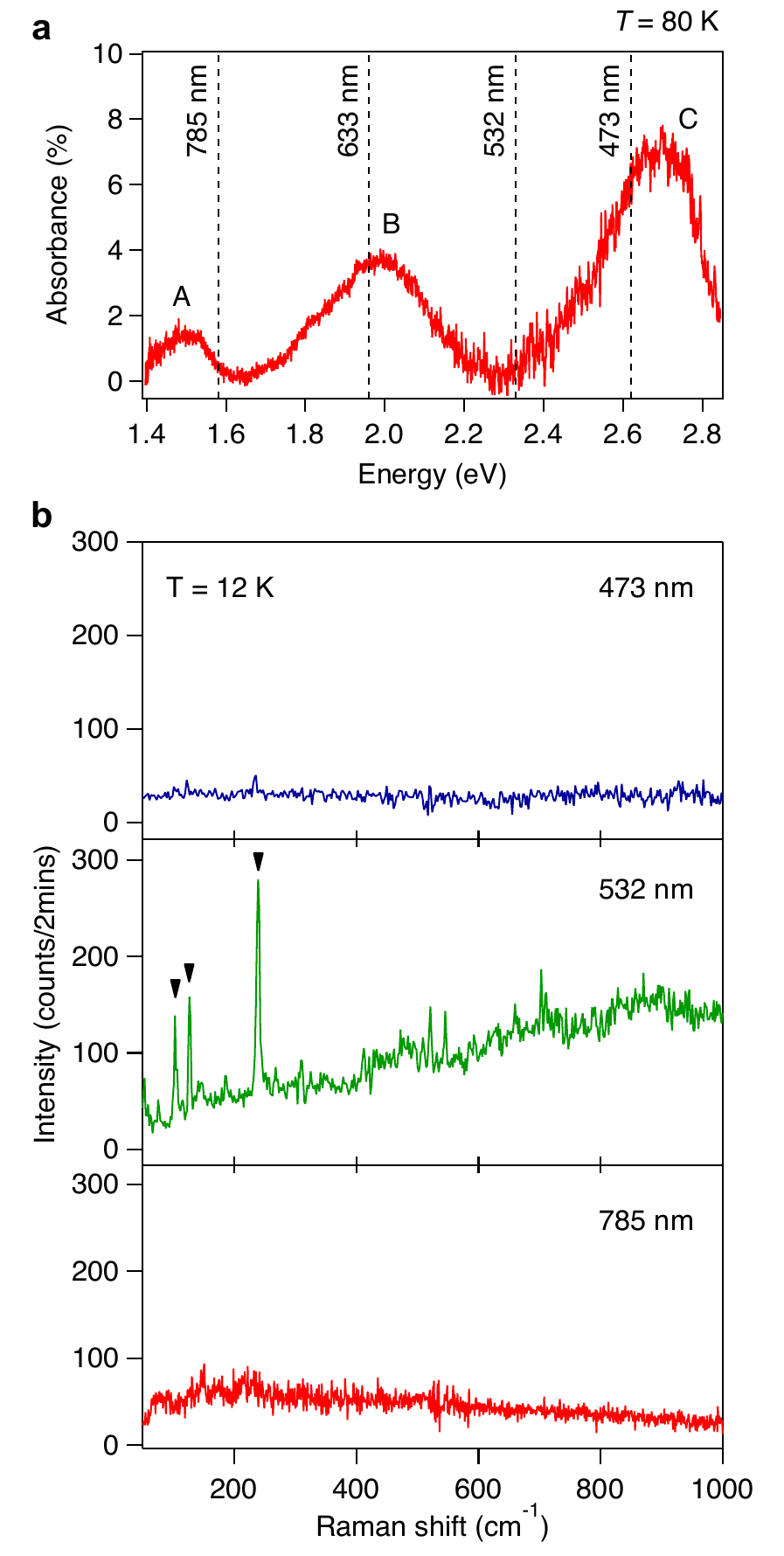}
\end{figure}
\begin{footnotesize}
\noindent \textbf{Fig. S5.} \textbf{a}. Absorption spectrum of bilayer \ce{CrI3} acquired at 80 K with the gradual background subtracted. \textbf{b}. Raman spectra taken on a bilayer \ce{CrI3} with three different incidence wavelengths: 473 nm, 532 nm, and 785 nm, in the linearly polarized parallel channel at 12 K. Black arrows in the 532 nm spectrum mark the first-order phonon modes in bilayer \ce{CrI3} that are consistent with the those phonons in the Raman spectrum taken by the 633 nm laser.\\
\end{footnotesize}

\noindent Scientifically, since the LO phonon involves in-plane atomic displacements between Cr and I atoms, making its coupling to the I 5$p$ to Cr 3$d$ charge transfer transition more efficient than any onsite transitions. Exciton A corresponds to $d$-$d$ transition within Cr 3$d$ orbitals, and therefore, it is expected that the LO phonon-A exciton coupling is weak. In contrast, B exciton corresponds to the I 5$p$ to Cr 3$d$ ($e_\mathrm{g}$ manifold) charge transfer resonance, and therefore couples to the LO phonon efficiently. Finally, C exciton is thought to be I 5$p$ to Cr 3$d$ ($t_\mathrm{2g}$ manifold), which in principle, should couple to the LO phonon as well. \\

\noindent Technically, 633 nm matches better with the B exciton than the other two wavelengths to A and C excitons. 785 nm only meets the upper tail of A exciton resonance and 473 nm is quite on the lower energy side of the C exciton resonance, whereas 633 nm is nearly at the B exciton resonance. This is consistent with the much weaker first-order phonon signals in the Raman spectra taken with 785 nm and 473 nm excitations than that with 633 nm excitation.\\

\noindent Further Raman studies with tunable CW lasers to precisely match A and C exciton resonances are needed to clarify if polaronic effect is present in these two types of excitons. \\

\newpage
\noindent \textbf{S6. Polaron Raman spectra of bilayer \ce{CrI3} in four circular polarization channels}\\

\noindent Figure S6 shows the Raman spectra of bilayer \ce{CrI3} in all four circular polarization channels. Above $B_\mathrm{C}$, (i) the ferromagnetic order has a net magnetization, and therefore the system responds differently to LL and RR (the two polarization geometries are related by the time-reversal operation); (ii) the ferromagnetic order has the inversion symmetry, and therefore Raman can only detect parity even modes.\\

\noindent Below $B_\mathrm{C}$, (i) the layered antiferromagnetic order has no net magnetization, and therefore the system responds equivalently to LL and RR (the two polarization geometries are related by the time-reversal operation); (ii) the layered antiferromagnetic order breaks inversion symmetry, and therefore it releases the constraint of only parity even modes being Raman active.\\

\begin{figure}[!h]
\includegraphics[scale=0.7]{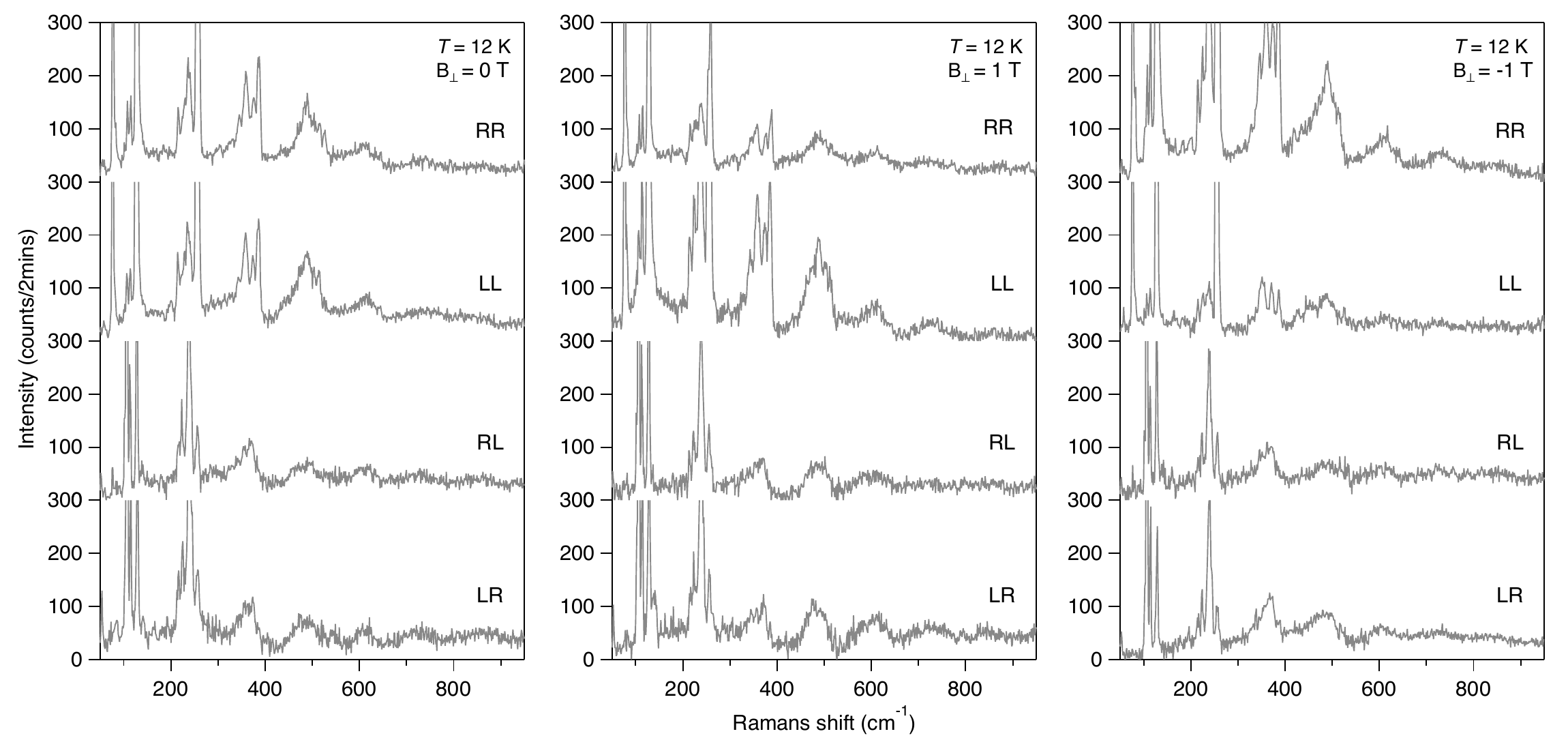}
\end{figure}
\begin{footnotesize}
\noindent \textbf{Fig. S6.} Raman spectra taken in all four circular polarization channels at 0 T (left), 1 T (middle), and -1 T (right). \end{footnotesize}

\newpage
\noindent \textbf{S7. Temperature and magnetic field dependent Raman measurements for tri-layer, four-layer, and five-layer \ce{CrI3}}\\

\noindent Figure S7 shows temperature dependent Raman spectroscopy results for tri-layer, four-layer, and five-layer \ce{CrI3}. We have performed the same analysis as we did for bilayer \ce{CrI3} to extract the peak intensity ($A_N$) v.s. $N^\mathrm{th}$ order for the periodic broad modes for the temperature range 10--65 K (with one spectrum every 5 K for all three thicknesses). We fitted $A_N$ with the Poisson distribution function to extract the electron-phonon coupling strength $\alpha_\mathrm{2D}$ for each temperature. We finally plotted $\alpha_\mathrm{2D}$ as a function of temperature. On a qualitative level, the temperature dependence results for these three thicknesses are consistent with that for 2L \ce{CrI3} in the main text Fig. 3.

\begin{figure}[!h]
\includegraphics[scale=0.55]{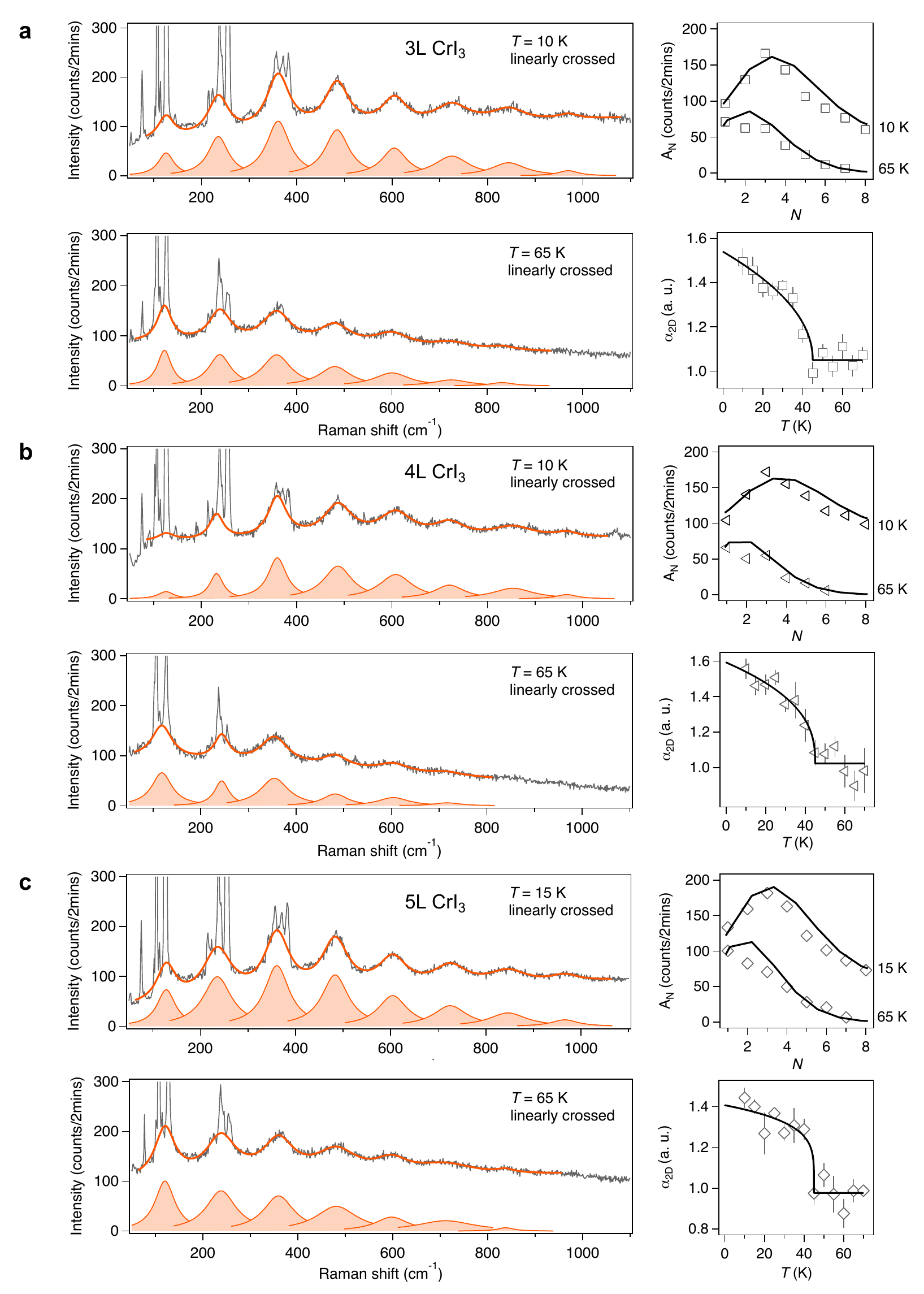}
\end{figure}
\begin{footnotesize}
\noindent \textbf{Fig. S7.} \textbf{a}. Temperature dependent Raman spectra taken on \textbf{a}. tri-layer (3L), \textbf{b}. four-layer (4L), and \textbf{c}. five-layer (5L) \ce{CrI3} with the same analysis as Fig. 3 in the main text. Error bars represent one standard error in the fits.\\
\end{footnotesize}

\end{document}